\def\be{\begin{equation}}
\def\ee{\end{equation}}
\def\bea{\begin{eqnarray}}
\def\eea{\end{eqnarray}}

\documentclass[final]{ias2}
\usepackage{bm}
\usepackage{graphicx}
\begin{document}
\title{Solitons and Matter Transport in Graphene Boundary}
\author[a]{Kumar Abhinav}
\email{kumarabhinav@iiserkol.ac.in}
\author[b]{Vivek M. Vyas}
\email{vivekmv@imsc.res.in}
\author[a]{Prasanta K. Panigrahi}
\email{pprasanta@iiserkol.ac.in}
\address[a]{Indian Institute of Science Education and Research (IISER) Kolkata, Mohanpur-741246, India}
\address[b]{Institute of Mathematical Sciences, Tharamani, Chennai-600113, India}

\begin{abstract}
It is shown that in 2+1 dimensional condensed matter systems, induced gravitational Chern-Simons (CS) 
action can play a crucial role for coherent matter transport in a finite geometry, provided zero-curvature 
condition is satisfied on the boundary. The role of the resultant KdV solitons is explicated. The fact that
KdV solitons can pass through each other without interference, can result in `resistance-less' energy
transport.
\end{abstract}

\maketitle

\section*{Introduction}
2+1 dimensional field theories naturally manifest in many physical systems, the prime example being
graphene \cite{1a,in1,1b}. Furthermore, the breaking of discrete parity and time-reversal symmetries can be realized in these
systems, by suitable mass-gap generation \cite{in2,in3}, which can produce a Chern-Simons (CS) action \cite{CS1,CS2}. Interestingly,
certain phonon modes like optical modes couple exactly like gauge fields \cite{in4}, making the realization of
CS theory a physicality. It has been found that, for finite size objects, these gauge theories can 
lead to novel form of superconductivity \cite{in5,in7}, with resistance-less boundary currents \cite{in7,in6}. Interestingly, some
of these systems carry defects, {\it e.g.}, pentagons besides hexagons in graphene, which can be described
by metric fields \cite{in8}, representing conic geometry. This raises the possibility of realizing 2+1 dimensional
gravity theories \cite{in9} in a condensed matter scenario.
\paragraph*{} In this paper, we demonstrate the intimate connection of CS theory, in a finite geometry, with
physics of 2+1 dimensional induced gravity and solitonic boundary excitations. It is found that,
energy transport can be realized in a manner analogous to resistance-less current in superconductors
\cite{in7,in6}. Induced gravity in graphene and its interaction with Dirac fermions will be discussed, followed by
a demonstration of natural emergence of KdV hierarchy in the boundary \cite{in9,in10,P2}.
\paragraph*{}The organization of the paper is as follows. We start with the example of U(1) gauge theory
realized in 2+1 dimensions with induced CS term, relevant for graphene, and show the implication
of finite geometry, in inducing resistance-less boundary currents. Then we will sow how the gravitational CS term 
emerge in graphene in an analogous manner. Subsequently, the gravitational analogy is
taken-up, wherein the boundary effects are related to solitonic physics.
In particular, Kdv solitons emerge as the boundary excitations, similar to the scenario
in case of Hall effect \cite{MS}.

\section{Induced CS Term in Graphene: Boundary Effects}
The hexagonal planar array of carbon atoms in graphene is resolvable into two inter-twined triangular sub-lattices.
The molecular orbitals between individual carbon atoms are such that the valence and conduction bands touch each other
at certain points in momentum space \cite{1a}. At low energies, near these touch-points, the dispersion is effectively 
linear and the characteristic tight-binding Hamiltonian, supporting nearest neighbor hopping, can be cast into a form
representing mass-less Dirac (relativistic) fermions in 2+1 dimensions, with speed of light ($c$) replaced by Fermi velocity $v_F$ \cite{in1}.
The experimental realization of this physics in the last decade \cite{1b}, re-initiated the study of such systems at 
theoretical, experimental and technological levels. As pure graphene sheet curls-up by its own, these systems were realized
as a single sheet of graphene placed on a suitable substrate that does not affect its electronic structure considerably.
Later-on, it was realized that opting for suitable substrate, like hexagonal boron nitrate (hBN), can break the sub-lattice 
symmetry of the system to induce massive Dirac excitations in graphene \cite{in2,in3}. 
\paragraph*{}2+1 dimensional quantum electrodynamics (QED) have been of much interest owing to certain advantages,
such as renormalizability \cite{Ren}, inherent topology and tractable infra-red divergences. For
massless fermions, it is infra-red finite \cite{Inf}. More interestingly the inherent topological
nature of such systems, automatically arising through quantum (loop) corrections, makes
these systems of great relevance to various physical phenomena, such as fractional Hall effect
\cite{QHE1,QHE2}, topological insulators \cite{TI1,TI2}, anyon superconductivity \cite{Any1,Any2} and many
more. With Schwinger pointing out that they can be massive \cite{Sch1}, construction of gauge theories
with non-zero mass arising through various mechanism became of interest. However, examples of intrinsically
massive tree (Classical) level Abelian gauge theories were proposed much later \cite{CS1,CS2}
in 2+1, owing to its intrinsic topological structure. Integrating out fermionic fields leads to an effective gauge theory
with gauge-invariant mass \cite{CS1,CS2,H} at low energies, through induction of the parity-violating topological CS term in
the action \cite{BDP}. In graphene, optical phonons have been shown to couple with Dirac fermions as gauge fields ($a_\mu$),
so as external photons ($A_\mu$) \cite{in4}. Due to the presence of two species of Dirac fermions in graphene \cite{in1},
owing to two sub-lattices of the hexagonal structure, the induced `pure' CS terms arising from either of these
fields mutually cancel-out. However, in the presence of both the gauge fields, the induced `mixed' CS terms survive,
leading to the effective Lagrangian, in the long-wavelength domain, as \cite{DP}, 

\bea
\mathcal{L}_{eff}&=&-\frac{1}{4\tilde{g}^2}f_{\mu\nu}f^{\mu\nu}-\frac{m}{2\pi\vert m\vert}\epsilon^{\mu\nu\lambda}A_{\mu}\partial_{\nu}a_{\lambda}-\frac{m}{2\pi\vert m\vert}\epsilon^{\mu\nu\lambda}a_{\mu}\partial_{\nu}A_{\lambda},\nonumber\\
f_{\mu\nu}&=&\partial_{\mu}a_{\nu}-\partial_\nu a_\mu.
\eea
Above form of Lagrangian is well-studied \cite{CS2,H,Wit}, and is appropriate for field-theoretic description of
graphene \cite{Wil}. It is further related to anyon superconductivity \cite{in7} and to fermions in 1+1 dimensions
through hydrodynamical analogy \cite{Sak}.
\paragraph*{}Consequently, the third term in $\mathcal{L}_{eff}$ is not gauge-invariant, resulting into a non-zero
change in action under U(1) gauge transformation:

\be
\delta S_{CS}=-\frac{m}{4\pi\vert m\vert}\int d^3x~\epsilon^{\mu\nu\rho}\partial_{\mu}\left(\Lambda F_{\nu\rho}\right),\nonumber
\ee 
$\Lambda$ being the local parameter of U(1) gauge transformation. Given the system has a finite boundary, 
the above expression yields a boundary integral as,

\be
\delta S_{B}=-\frac{m}{4\pi\vert m\vert}\oint_B d^2x~\epsilon^{\mu\nu}\Lambda F_{\mu\nu}.\label{v1}
\ee
This term can also result from a pure CS Lagrangian, modulo a constant factor, subjected to {\it large} gauge transformations
that leads to a non-zero winding number. As will be shown later, this property enables us to link the above system with $SL(2,R)$
KdV solitons.
\paragraph*{}The consistency with the gauge-invariance at the tree-level
demands the existence of a compensating term at the boundary, allowed by the variable redundancy 
of the gauge theory. The simplest choice is \cite{in7},

\be
S'_{B}=\oint_B d^2x\left[\frac{c}{2}\left(\partial_\mu\theta-A_\mu\right)^2+\frac{m}{4\pi\vert m\vert}\theta\epsilon^{\mu\nu}F_{\mu\nu}\right].\nonumber
\ee 
The St\"uckelberg field \cite{u1} $\theta$ transforms as $\theta\rightarrow\theta+\Lambda$ under gauge
transformation, rendering it to be massless for overall gauge-invariance. The above integral is gauge-invariant
only on the mass-shell, yielding the equation of motion:

\be
\partial^2\theta=\frac{m}{4\pi c\vert m\vert}\epsilon^{\mu\nu}F_{\mu\nu},\label{a}
\ee 
The above result is of prime importance in the present work, as it shows dependence of $\theta$ on
the field-strength tensor. As will be shown in the next section, the field-strength tensor is
equivalent to the curvature tensor for a non-trivial geometry, with spin connections serving as
components of the equivalent gauge field \cite{g4}. The corresponding gravitational action, in 2+1 dimensions,
is also of the CS nature \cite{g6}, yielding a surface term of the form in Eq. \ref{v1}. Therefore, 
the present theory equivalently describes non-trivial gravitational excitations at the graphene boundary,
physically realizable through defects \cite{in8}. Further, as the resultant geometry is conic in nature,
it leads to a metric diagonal in light-cone coordinates, having $SL(2,R)$ symmetry that supports solitonic
excitations of KdV form under the zero-curvature condition. The $\theta$ field will be identified with these
solitons.
\paragraph*{}Eq. \ref{a} is dual to the Schwinger model \cite{u2}, incorporating massless fermions, coupled to a gauge
field with coupling constant $1/\sqrt{4\pi c}$. For such a theory, defined on a circle, the chiral current 
is anomalous: 

\be
\partial_\mu j_5^\mu\propto\epsilon^{\mu\nu}F_{\mu\nu},~~~F_{\mu\nu}=\partial_{\mu}A_{\nu}-\partial_\nu A_\mu,
\ee
Therefore, the graphene edge states are chiral and gapless. On considering gauge transformation for the 
external field too, we end up with the anomaly equation,

\be
\partial_\mu j_5^\mu\propto\epsilon^{\mu\nu}\left[F_{\mu\nu}+f_{\mu\nu}\right].
\ee
In 2+1 dimensions, such a theory obtained through integrating-out four-component Dirac spinor fields,
is known to support unconventional decay modes of lattice vibrations \cite{PKPz}.
\paragraph*{}Therefore, though the bulk action is both parity and time-reversal invariant, the surface
one is not. All these discussions hold for another effective Lagrangian,

\be
\mathcal{L}_{eff}=-\frac{1}{6\pi\vert m\vert}f_{\mu\nu}f^{\mu\nu}-\frac{m}{\pi\vert m\vert}\epsilon^{\mu\nu\lambda}A_{\mu}\partial_{\nu}a_{\lambda}\nonumber
\ee
\paragraph*{}Again, considering $\mathcal{L}_{eff}$ and integrating out the $a_{\mu}$ field, yields,

\be
\mathcal{L}_{eff}[A]=-\frac{\tilde{g}^2}{4\pi^2}F_{\mu\nu}\frac{1}{\partial_\rho\partial^\rho}F^{\mu\nu}\nonumber
\ee
In the Lorentz (covariant) gauge, the corresponding induced electric current is found to be,

\be
\left\langle\vec{j}(x)\right\rangle=-\frac{\tilde{g}^2}{\pi^2}\vec{A}(x),
\ee
ensuring superconductivity. However, one can see that the action for the $\theta$
field, $S'_{B}$ also is in the manifest London form \cite{u3}. Therefore, both bulk and the boundary
supports resistance-less conductivity. However, for the boundary case, it is inherently of 
topological origin and arises through anomalous fermionic current. Further, it is seen to 
support only the combination $a_{\mu}+A_{\mu}$, not $a_{\mu}-A_{\mu}$. Therefore the resistance-less
transport is supported by the arm-chair edge, not the zig-zag edge, of graphene \cite{in7}.
\paragraph*{}It is further evident that the massless chiral mode $\theta$, also corresponds
to resistance-less matter transport on the boundary, owing to its equivalence to KdV solitons,
which will be explicated later. 

\section{Emergent Gravity in Graphene}
Defects can introduce an equivalent gravitational background in graphene in many
ways \cite{u4}, as Dirac fermions move in locally curved space, making it analogous
to 2+1 dimensional quantum gravity \cite{u5}. More specifically, replacement of hexagonal structure
with pentagon (heptagon) results into positive (negative) curvature \cite{g1,g2,u6}, which
can be interpreted equivalently as one created by a point particle under which the Dirac fermions
are moving \cite{g3,g4,g5}. Effectively, such a scenario is valid at a distance from the defect large
compared to its own dimensions, as the effective Dirac dispersion is invalid near such a defect owing 
to the absence of hexagonal symmetry. 
\paragraph*{}The motion of a Dirac particle in the gravitational field due to a point particle has been
studied by introducing the metric \cite{g4},

\be
ds^2=dt^2-\frac{1}{\beta^2}dr^2-r^2d\theta^2,~~~\beta:=1-4GM,\nonumber
\ee
with $M$ being the effective mass of the point particle, which can be fixed from the defect-induced
curvature of graphene itself.  
\paragraph*{}Motion of a particle, having mass and spin in general, in a space of non-trivial metric
is a well-studied problem \cite{D}. At the equation of motion level, the ordinary derivative gets extended
to co-variant derivative, including gravitational couplings to mass and spin through affine
and spin connections. The general picture is best presented through introducing a flat coordinate
system at each point in space, and a generic one with curvature incorporated, with degrees of freedom
respectively denoted by Latin and Greek indices. The connection between the two is provided by the
`vielbein' tensors, $e^a_\mu$, with $e^a_\mu e^{\mu b}:=\eta^{ab}$ and $e_a^\mu e^{a\nu}:=g^{\mu\nu}$
being the local `flat' and general `curved' metric tensors respectively. All the other tensors can be
defined accordingly. 
\paragraph*{}In the case of interest, the induced curvature is conic, and there, the massive Dirac Lagrangian
takes the form \cite{g4}:

\bea
\mathcal{L}_D&=&\bar{\psi}(x)\left(i\gamma^ae^{~\mu}_aD_\mu-m\right)\psi(x),\nonumber\\
D_\mu&=&\partial_\mu+\frac{1}{2}\omega_{\mu,ab}\sigma^{ab},~~~\sigma^{ab}:=\frac{1}{4}[\gamma^a,\gamma^b].\label{G1}
\eea
where $\gamma^a$'s are 2+1 Minkowski-space Dirac matrices that transform under local Lorentz transformations
incorporating curvature, $\omega_{\mu,ab}$ are spin-connections and $e^{~\mu}_a$ is the {\it vielbein} for
the metric given above. Further, as,

\be
\omega_{\mu,ab}=\epsilon_{abc}\omega_\mu^c~~~\text{and}~~~\sigma^{ab}=-\frac{i}{2}\epsilon^{abc}\gamma_c,\nonumber
\ee
the covariant derivative becomes $D_\mu=\partial_\mu-\frac{i}{2}\omega_\mu^{~a}\gamma_a$.
Thus, $-\frac{c}{2e}\omega_\mu^{~a}\gamma_a$ serves the purpose of a gauge field, and Eq. \ref{G1}
can lead to the corresponding CS gauge theory. 
\paragraph*{}In the above  frame-work, the Einstein-Hilbert action, on a manifold $M$ can be written
as \cite{g6}:

\be
I:=\frac{1}{2}\int_M\epsilon^{\mu\nu\rho\lambda}\epsilon_{abcd}e^{~a}_\mu e^{~b}_\nu R_{\rho\lambda}^{~~~cd},\nonumber
\ee
with the curvature tensor $R_{\rho\lambda}^{~~~cd}(\omega)$, defined in terms of spin-connections 
$\omega_\mu^{~a}$, is related to the usual Riemann tensor $R^\mu_{\nu\rho\lambda}(\Gamma)$, defined 
in terms of affine connections $\Gamma^\mu_{\nu\rho}$, as:

\be
R^\alpha_{\lambda\mu\nu}(\Gamma)=e^{~\alpha}_ae_{\lambda~b}R^{ab}_{~~\mu\nu}(\omega),~~~R^{ab}_{~~\mu\nu}(\omega):=\partial_\mu\omega^{~a}_{\nu~b}-\partial_\nu\omega^{~a}_{\mu~b}+[\omega_\mu,\omega_\nu]^a_{~b}.\nonumber
\ee 
Variation with respect to vielbeins lead to the flat-space (vacuum) Einstein equation:

\be
e^\mu_{~a}R_{\mu\nu~~b}^{~~~a}=0,\nonumber
\ee
equivalent to the vanishing of the Ricci tensor $R_{\mu\nu}(\Gamma)=e^{\nu}_{~b}e^{\alpha}_{~b}R_{\mu\alpha~~b}^{~~~a}(\omega)$,
representing {\it zero-curvature} condition, which will lead to KdV hierarchy, to be shown
in the next section. Further, on can readily identify the vielbein and spin connections as
`gauge' fields here, which makes the curvature tensor effectively the gravitational field-strength
tensor corresponding to these gauge fields, more specifically, that of the spin connection.
\paragraph*{}In 2+1 dimensions, the Einstein-Hilbert action takes the form,

\be
I=\frac{1}{2}\int_M\epsilon^{\mu\nu\rho}\epsilon_{abc}e^{~a}_\mu R_{\nu\rho}^{~~~bc},\nonumber
\ee
which Witten \cite{g6} showed to be a CS action which can be written as:

\be
I_{CS}=\int_M\epsilon^{\mu\nu\rho}e_{\mu a}R^{~~~a}_{\nu\rho},~~~R^{~~~a}_{\nu\rho}:=\partial_\nu\omega_\rho^{~a}-\partial_\rho\omega_\nu^{~a}+\epsilon^{abc}\omega_{\nu b}\omega_{\rho c},\label{w1}
\ee
with curvature tensor $R^{~~~a}_{\alpha\beta}$ serving as field strength tensor corresponding to
spin connections as gauge fields. It is equivalent to the `usual' non-Abelian CS action:

\be
I'_{CS}=\int_M\epsilon^{\mu\nu\rho}\omega_{\mu a}\left[\partial_\nu\omega_\rho^{~a}-\partial_\rho\omega_\nu^{~a}+\frac{2}{3}\epsilon^{abc}\omega_{\nu b}\omega_{\rho c}\right].\label{w2}
\ee
both yielding the same equation of motion, 

\be
\epsilon^{\mu\nu\rho}R^{~~~a}_{\nu\rho}=0\label{eom}.
\ee
corresponding to the variation with respect to $e_\mu^a$ and $\omega_\mu^{~a}$ respectively, corresponding
to zero-curvature condition as mentioned earlier. The  local Lorentz (gauge)
transformations for vielbein and spin connection are \cite{Fu}:
\bea
\delta e^{\mu}_k&=&-\Lambda_k^{~n}e^{\mu}_n,\nonumber\\
\delta\omega_\mu^{~a}&=&\partial_\mu\Lambda^a,~~~\Lambda^a:=\frac{1}{2}\epsilon^{abc}\Lambda_{bc}\nonumber
\eea
where $\Lambda_{ab}$ is the parameter of transformation. Under these, $I_{CS}$ and $I'_{CS}$ respectively behave as mixed and pure CS
actions on the boundary as:

\bea
\delta I_{CS}&=&-\int_M e_{\mu b}\epsilon^{\mu\nu\rho}\Lambda_a^{~b}R^{~~~a}_{\nu\rho}\equiv -\int_Y\epsilon^{\mu\nu}\Lambda_aR^{~~~a}_{\mu\nu},\label{w3a}\\
\delta I'_{CS}&\propto&\int_M \epsilon^{\mu\nu\rho}\epsilon^{abc}\partial_\mu\Lambda_a\partial_\nu\Lambda_b\partial\rho\Lambda_c\equiv\int_Y\epsilon^{\mu\nu}\epsilon^{abc}\Lambda_a\partial_\nu\Lambda_b\partial\rho\Lambda_c\label{w3b}
\eea  
A few points here are of importance. In Eq. \ref{w3a}, by choosing the vielbein $e_\mu^{~a}$ as a unit normal to the
1+1 dimensional hypersurface $Y$ bounding the 2+1 dimensional manifold $M$, we get a `surface-term' on $Y$ similar to that in Eq. \ref{v1}.
In Eq. \ref{w3b}, the first integrand is equivalent to that corresponding to the winding number for non-Abelian
CS theory for large gauge transformations in the $SU(2)$ case, $\epsilon^{\mu\nu\rho}T_r\left[\partial_\mu UU^{-1}\partial_\nu UU^{-1}\partial_\rho UU^{-1}\right]$, with
$U:=\exp\left(i\Lambda_aT^a\right)$ and gauge-group generators satisfying $Tr\left[T^aT^bT^c\right]=\epsilon^{abc}$,
also leading to a surface-term on $Y$. Therefore, the equivalence of $I_{CS}$ and $I'_{CS}$ leads to the interpretation
of the compensation condition (Eq. \ref{a}), for the gauge-variant topological boundary term in graphene.
\paragraph*{}The non-zero boundary integral is of pure topological origin and is unaltered by the presence of other 
tree-level terms in the gauge Lagrangian. This is reflected in similarity of Eqs. \ref{w3a} and \ref{v1}. In order to
restore gauge-invariance at the boundary, similar introduction of a `generalized' St\"uckelberg matrix-valued field
$\theta^a$ leads to,

\be
\partial^2\theta^a\propto\epsilon^{\mu\nu}R^a_{\mu\nu},\label{v3}
\ee
with the curvature tensor $R^a_{\mu\nu}$ replacing the field-strength tensor. Therefore, the resistance-less
massless chiral mode $\theta$, obtained in Section 1, has an gravitational equivalent owing to gravity being
a CS gauge theory in 2+1 dimensions. The earlier being a scalar while the later being a Lorentz vector is 
justified as at the 1+1 dimensional boundary, only one component of the gravitational field stay dynamic, owing 
to the zero-curvature condition. The physical origin of the gravitational `gauge' field in graphene can 
be the defect-induced curvature. The fact that the same is conic in nature, enables light-cone metric formulation
at the 1+1 dimensional boundary. Further, as the zero-curvature condition is automatically satisfied at the 
equation of motion level, the KdV hierarchy naturally emerge from this, as will be shown in the next section.   

\section{KdV hierarchy from Zero Curvature condition in 1+1}
As seen above, boundary gravity in 2+1 dimensions arises through interaction of matter with gravitational field.
In the light-cone gauge, this theory can be shown to incorporate $SL(2,{\mathbb R})$ current algebra \cite{in9,in10},
enabling a representation of the theory in terms of partial differential equations. The light-cone geodesic is given by,

\be
ds^2=dx^+dx^-+g_{++}(x^+,x^-)\left(dx^+\right)^2,\nonumber
\ee
with scalar curvature,

\be 
R=\partial_-^2g_{++},~~~R=e_a^{~\mu}e_b^{~\nu}R_{\mu\nu}^{~~~ab}(\omega).\label{R}
\ee
The Ward identities for the metric assumes the form \cite{in9}:

\be
\kappa\frac{\partial}{\partial x_j^+}\left\langle f(x_1)...f(x_N)\right\rangle=\sum_{k\neq i}\frac{\eta_{ab}l_i^al_k^b}{x_i^+-x_k^+}\left\langle f(x_1)...f(x_N)\right\rangle
\ee
where $\kappa$ is arbitrary, $\eta_{ab}$ is the Killing metric tensor for $SL(2,R)$ and $f$ is
defined through: $\partial_+f=g_{++}\partial_-f$. $l_j^a$'s are identified as the $SL(2,{\mathbb R})$ generators.
From this differential equation interpretation
of 2-D gravity, it is possible to obtain a KdV hierarchy \cite{in10} under zero-curvature 
constraint, as well as a Lax pair formalism \cite{P2}, both supporting solitonic excitations.
\paragraph*{}The zero-curvature condition physically corresponds to path-independence of evolution
of a physical state in a generic space-time. In terms of covariant derivative $D_\mu$, defined on a 
non-trivial manifold, the curvature tensor is defined as the commutator $[D_\mu,D_\nu]$. The vanishing
of the same means parallel transport around any closed loop being identity. Over a non-trivial gauge
transformation, a complete cycle around the loop results into the Wilson loop operator
$W(\lambda)=\mathcal{P}\exp\left[i\int dx_\mu A^{\mu}(x;\lambda)\right]$ \cite{wil} with the end points
identified. The factor $\lambda$ parametrizes the loop. For the zero-curvature condition been satisfied,
the asymptotic conditions $\lambda\rightarrow\infty,0$ yields, respectively,

\be
\log Tr W(\lambda)\rightarrow\lambda^{\pm 1} R-\sum_{k=1}^\infty\lambda^{\pm k}H^\pm_k,~~~H_k^\pm=i\int dx_\mu\left(I^\pm_k\right)^\mu,\nonumber
\ee
with infinite number of conserved quantities $H_k^\pm$. This renders the system to be integrable \cite{01}.
Therefore, a system incorporating a physical notion of curvature can always be constrained by imposing
the zero-curvature condition to yield an integrable model. In case of 2-D gravity, the same leads to the
KdV hierarchy \cite{in10}.
\paragraph*{}A 1+1 dimensional $SL(2,R)$ Lie algebra, in light-cone coordinates, can be represented by
the corresponding Lie-algebra-valued non-Abelian gauge fields $A_\pm=A_\pm^al_a$ in a $2\times2$
representation. $A_-$ is suitably chosen and hence partially gauge-fixed. Subsequently,
the only independent component of the field strength tensor in 1+1 is,

\bea
F_{+-}&=&\partial_+A_--\partial_-A_++[A_+,A_-]\nonumber\\
&=&\left( \begin{array}{cc} 0 & -\partial_+u+\frac{1}{2}\partial^3_-C+2u\partial_-C+C\partial_-u-2\lambda^2\partial_-C \\ 0 & 0 \end{array} \right),\label{b}
\eea
where $\lambda$ is a global constant (spectral parameter), $u=u(x_+,x_-)$ and $C=C(u,\lambda)$. 

\paragraph*{}As discussed above, the evolution of the KdV integrable system, can be interpreted as a zero-curvature
condition \cite{01} associated with some gauge group. More specifically, the KdV hierarchy can be obtained 
from the $SL(2,R)$ zero curvature condition, which can be interpreted as gauge anomaly equation \cite{in10}
and WZWN action \cite{WZWN,02}. The $SL(2,R)$ Lie-group generator $U(x)=\exp\left(i\Theta^a(x)l_a\right)$
results into a matrix valued vector $A_\mu(x):=U^{-1}(x)\partial_\mu U(x)$ in 1+1 dimension (pure gauge). This
automatically yields the `zero curvature' condition: $F_{\mu\nu}:=\partial_\mu A_\nu-\partial_\nu A_\mu+\left[A_\mu,A_\nu\right]=0$
Further, this condition makes path-ordered integral with $F_{\mu\nu}$ exponentiated to be path-independent,
depending only on the end-points \cite{01}, bringing-out the topological nature of the system.
\paragraph*{} Presently, the zero-curvature condition in case of $SL(2,R)$ Lie
group, {\it i. e.}, $F_{+-}=0$ leads to the $n$-th order KdV hierarchy equation,

\be
\partial_+u=\frac{1}{2}\left[\partial^3_-+2(\partial_-u+u\partial_-)\right]C_n.\label{c}
\ee
on Taylor-expanding $C(u,\lambda)$ in powers of spectral parameter $\lambda$. For the special case of $\lambda=0$,
infinitesimal gauge transformation leads $C$ to mirror the light-cone metric tensor transformation under infinitesimal conformal
transformation \cite{k7}, identifying,

\be
C=g_{++}.
\ee
\paragraph*{}Therefore, the non-trivial metric component results into KdV hierarchy, enabling solitonic excitations
in the 1+1 space, akin to Hall effect. Considering the manifold to be the boundary of a 2+1 space,
we may realize the same at the graphene boundary, yielding,

\be
\partial^2\theta=0=\partial^2\theta^a,
\ee
which is equivalent to $\partial^2_-C=0$ (Eq. \ref{R}), thereby leading to the inference that $\theta$ represent 
a solitonic excitation. It is to be noted that although the original CS gauge theory in graphene was Abelian in
nature, the equivalent induced conic gravity, in presence of `suitable' defects, yields a non-Abelian counterpart.
The later supports 1+1 dimensional boundary modes that yields resistance-less transport of solitonic matter subjected
to the zero-curvature condition. Intuitively, the path-independence of the gravitational system reflected through
the zero-curvature condition directly corresponds to the topological nature of the CS gauge theory.
\paragraph*{}Therefore, $\theta^a$ represents a mass-less mode at the arm-chair boundary, which is decoupled from the curvature (induced gravity)
within the bulk. This mode is chiral in nature, owing to its CS origin and has an associated solitonic excitation
owing to the emergent KdV hierarchy condition (Eq. \ref{c}). Therefore, given the zero-curvature 
condition is satisfied, the anomalous topological boundary term supports `resistance-less' solitonic energy
transport at graphene boundary. This mode may experimentally be realized at the arm-chair edge. 

\section*{Conclusions}
In conclusion, matter induced gravitational Chern-Simons term can be realized in condensed matter
systems like graphene when defects are present. In a system with a finite geometry, the induced CS
term is shown to yield energy transport through solitons under zero-curvature condition at the boundary.
This is akin to manifestation of solitonic boundary excitations in Hall effect. The situation with non-zero
boundary curvature will be discussed elsewhere.

\end{document}